\title{Cosmic Background Removal with Deep Neural Networks in SBND} 

\date{\today}

\documentclass[11pt]{article}

\usepackage{url,lineno,microtype,subcaption, graphicx}
\usepackage[margin=0.5in]{geometry}
\usepackage[colorlinks=true]{hyperref}
\usepackage{authblk}

\author[15]{R.~Acciarri}
\author[1]{C.~Adams}
\author[21,29]{C.~Andreopoulos}
\author[34]{J.~Asaadi}
\author[6]{M.~Babicz}
\author[35]{C.~Backhouse}
\author[15]{W.~Badgett}
\author[15]{L.~Bagby}
\author[30]{D.~Barker}
\author[23]{V.~Basque}
% ...

\affil[1]{Argonne National Laboratory, Lemont, IL 60439, USA}
% ...
\affil[6]{CERN, European Organization for Nuclear Research 1211 Geneve 23, Switzerland, CERN}
% ...
\affil[15]{Fermi National Accelerator Laboratory, Batavia, IL 60510, USA}
% ...
\affil[21]{University of Liverpool, Liverpool L69 7ZE, United Kingdom}
% ...
\affil[23]{University of Manchester, Manchester M13 9PL, United Kingdom}
% ...
\affil[29]{STFC, Rutherford Appleton Laboratory, Harwell OX11 0QX, United Kingdom}
\affil[30]{University of Sheffield, Department of Physics and Astronomy, Sheffield S3 7RH, United Kingdom}
% ...
\affil[34]{University of Texas at Arlington, TX 76019, USA}
\affil[35]{University College London, London WC1E 6BT, United Kingdom}
% ...

\begin{document}
\onecolumn

% \date{\today}% It is always \today, today,
%              %  but any date may be explicitly specified
\maketitle

\begin{abstract}
In liquid argon time projection chambers exposed to neutrino beams and running on or near surface levels, cosmic muons and other cosmic particles are incident on the detectors while a single neutrino-induced event is being recorded.
In practice, this means that data from surface liquid argon time projection chambers will be dominated by cosmic particles, both as a source of event triggers and as the majority of the particle count in true neutrino-triggered events.
In this work, we demonstrate a novel application of deep learning techniques to remove these background particles by applying semantic segmentation on full detector images from the SBND detector, the near detector in the Fermilab Short-Baseline Neutrino Program. We use this technique to identify, at single image-pixel level, whether recorded activity originated from cosmic particles or neutrino interactions.

\end{abstract}

%\tableofcontents

\section{\label{sec:Intro}Introduction}

Liquid argon time projection chambers (LArTPCs) are high resolution, calorimetric imaging particle detectors.  Due to their excellent calorimetric properties and particle identification capabilities \cite{argoneut_nue}, combined with their scalability to kiloton masses \cite{dune-tdr-1}, LArTPCs have been selected for a variety of experiments to detect neutrinos in the MeV to GeV energy range.  
Several 100 to 1000-ton-scale LArTPCs have collected substantial amounts of neutrino data (ICARUS at LNGS \cite{Rubbia_2011} and MicroBooNE at Fermilab \cite{uboone_detector}), or been operated in charged particle test beams (ProtoDUNE-SP \cite{protodune-singlephase-tdr} and ProtoDUNE-DP \cite{dune-tdr-3} at CERN). Others are in the commissioning phase (ICARUS at Fermilab \cite{sbn-proposal}) or under construction (SBND at Fermilab \cite{sbn-proposal}).  Coming later this decade, the Deep Underground Neutrino Experiment, DUNE \cite{dune-tdr-2}, will be a $10^4$-ton-scale LArTPC neutrino detector built 1.5 km underground in the Homestake Mine in South Dakota.

LArTPCs running near the Earth’s surface (such as SBND, MicroBooNE, and ICARUS comprising the Short-Baseline Neutrino (SBN) program at Fermilab) are susceptible to backgrounds induced by cosmic interactions, which occur at much higher rates than neutrino interactions.
In this paper, we present novel techniques for the tagging of cosmic-induced, neutrino-induced, and background-noise pixels, using deep learning and image processing techniques applied to simulated data from the SBND LArTPC detector. 

We first present, in Section~\ref{sec:sbnd}, a description of the liquid argon time projection chamber technology, particularly in the context of the SBND experiment where this study is performed.  In Section~\ref{sec:problem} we summarize the origin of the problem we solve with convolutional neural networks, including a description of how LArTPC images are created from the raw data for this study.  Section~\ref{sec:related} summarizes the related work on this challenge, and Section~\ref{sec:dataset} describes the details of the dataset used in this study.  Sections~\ref{sec:network} and \ref{sec:training} describe the design and training of the convolutional neural network, respectively, and Section~\ref{sec:analysis} presents a basic analysis based on the trained network.

\section{The SBND Liquid Argon Time Projection Chamber}
\label{sec:sbnd}

The LArTPC is a high resolution, high granularity, scalable particle detector.  Many detailed descriptions of LArTPCs are available \cite{icarus_detector, uboone_detector, argoneut} but we will summarize the key features here.  In this discussion, we will focus on the near detector of the SBN Program at Fermilab, the Short Baseline Near Detector or SBND, since it is the origin of the dataset used here.

%% the below includes edits of the commented out text below (DS)
A LArTPC is an instrumented volume of purified liquid argon under an approximately uniform electric field.   At one side is the source of the electric field, the cathode.  At the other side, the anode, are readout channels to detect charge.  In SBND, the readout channels are wire-based.  

When charged particles traverse the active argon region, they ionize the argon atoms and leave a trail of argon ions and freed electrons.  The freed electrons drift under the influence of the electric field toward the sense wires, where they are detected either via induction or directly collected on the sense wires.  Each wire is digitized continuously, and the time of charge arrival indicates how far the charge drifted.  A very thorough description of the mechanisms and signal processing for wire-based TPCs can be found in \cite{uboone_sigproc_1, uboone_sigproc_2}. 

\begin{figure}
    \centering
    \includegraphics[width=\textwidth]{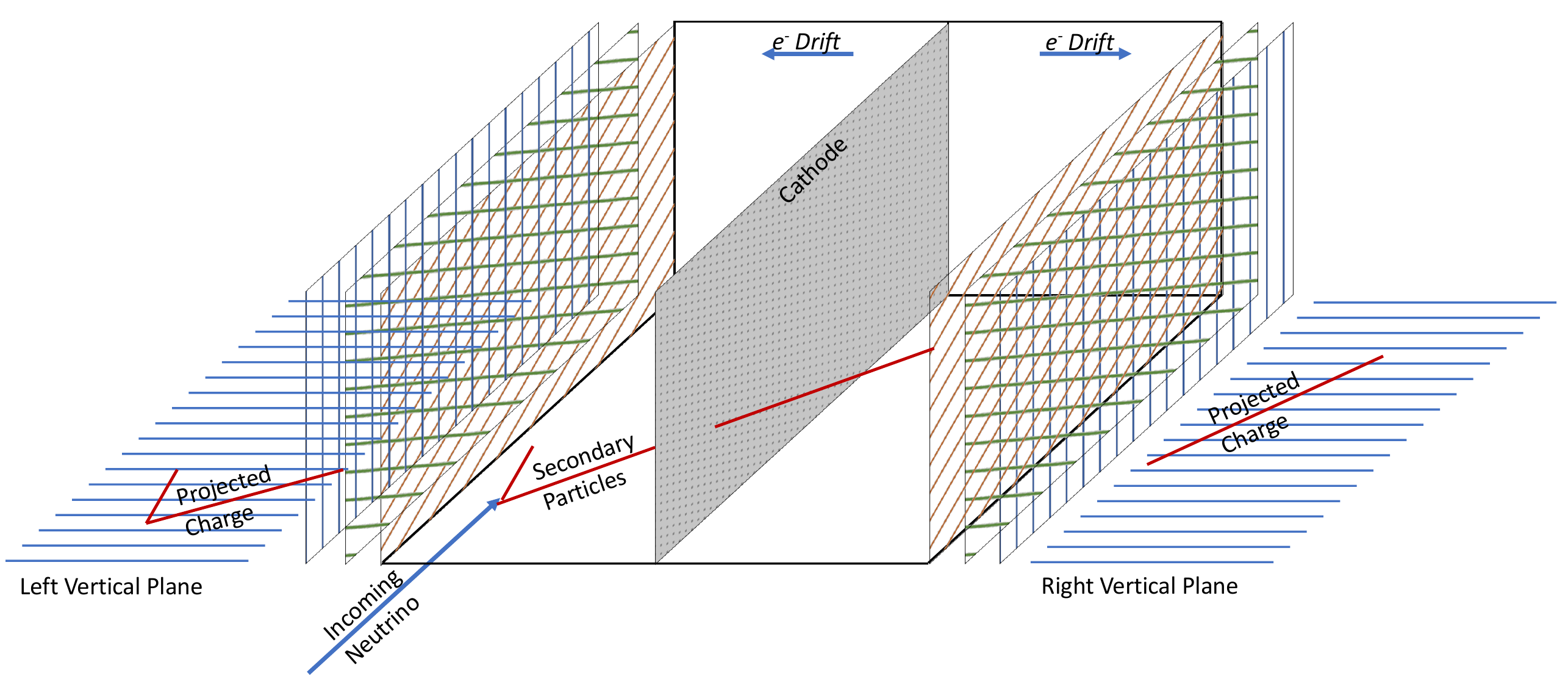}
    \caption{An illustration of the SBND TPC used in this work.  In this image, a neutrino interacts in the left TPC, and the outgoing particles cross the central cathode into the right TPC.  The top-down projection images (vertical wire planes) are shown, which are combined into one image as seen in Figure~\ref{fig:raw_data}.}
    \label{fig:lartpc_schematic}
\end{figure}

\begin{figure}
    \centering
    \includegraphics[width=0.4\textwidth]{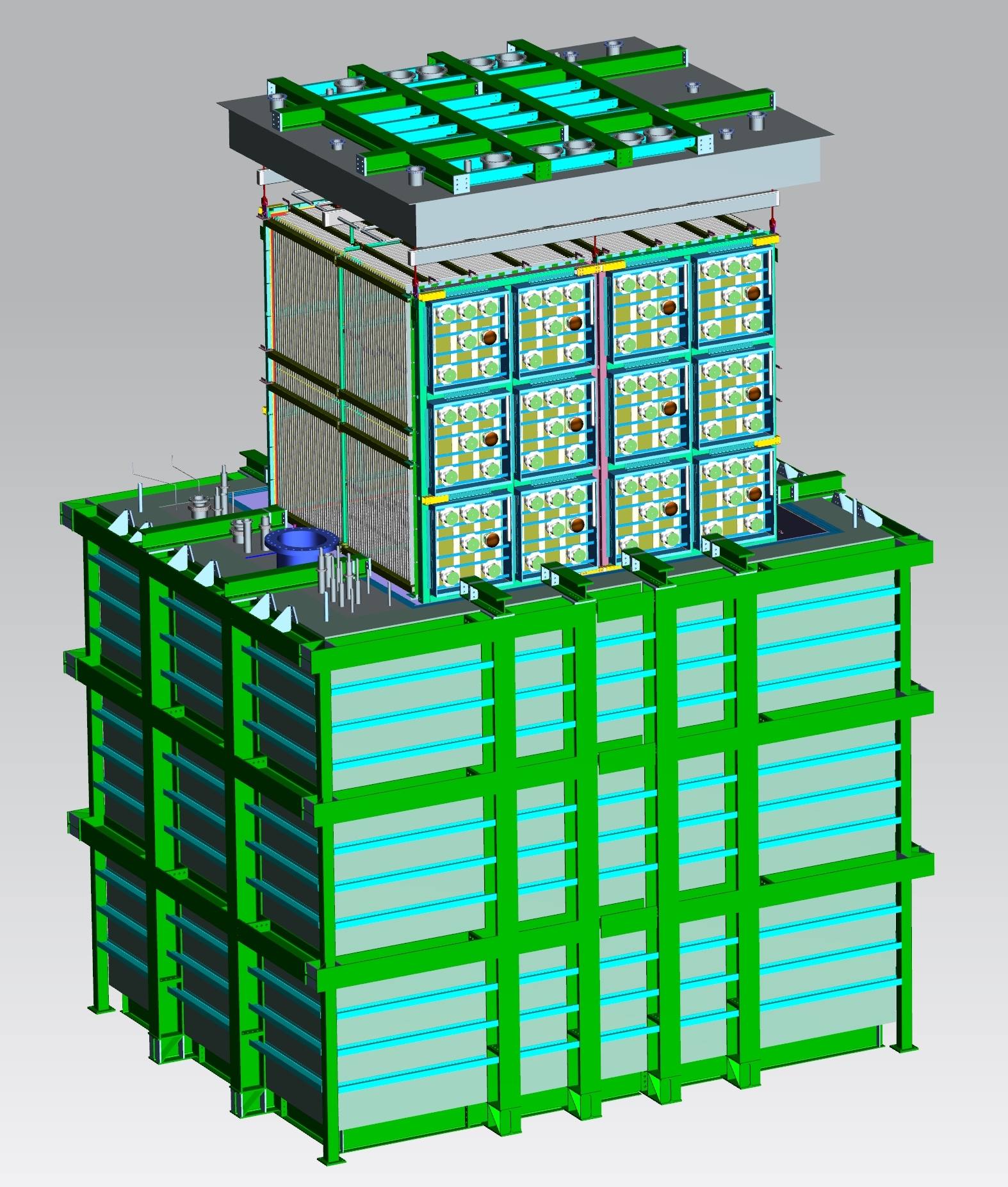}
    \caption{Engineering diagram of the SBND LArTPC and its surrounding subsystems.  Here, the TPC is shown lifted above the cryostat for clarity.}
    \label{fig:sbnd}
\end{figure}

The SBND detector is a dual drift TPC, with a central, shared cathode and two anodes, one at each side of the detector (see Figure~\ref{fig:lartpc_schematic}).  The vertical wire planes each have 1664 wires (plane 2 in images in this work), and each of the induction planes (angled at +/- 60 degrees, planes 0 and 1 in this work) have 1984 wires \cite{apa_paper}.  Each TPC is approximately 5 meters long, 4 meters high, and 2 meters in the drift direction - for a total width of approximately 4 meters.  The entire TPC is located within a cryogenic system, as seen in Figure~\ref{fig:sbnd}.

SBND is also surrounded, nearly entirely, by a solid scintillator-based cosmic-ray muon tracking (CRT) system.  The CRT observes the passing of cosmic muons and provides their time of arrival, in principle allowing a veto of some cosmic ray interactions that have no neutrino interactions.  Additionally, the interior of the LArTPC detector has a photon detection system to collect the prompt scintillation light that is also generated by charged particles traversing the argon.  Both the CRT and photon collection systems could be useful for disentangling cosmic-only and cosmic-with-neutrino events (as described in Section~\ref{sec:problem}), but in this work we focus exclusively on analysis of TPC data in the form of 2D images.

SBND is located in the Booster Neutrino Beam at Fermilab, and will observe neutrino interactions in an energy range from a few hundred MeV to several GeV. The SBND detector is under construction at the time of this writing, and results here use simulations based on the design of the detector.

\section{Problem Description}
\label{sec:problem}
We seek in this work to remove background activity generated by cosmic particle interactions in the SBND dataset, and in this section we will describe in more detail how the SBND LArTPC operates and why cosmic interactions are problematic.

During typical operation, a LArTPC digitizes the entire detector for a period of time, usually equal to or larger than the time needed for an ionization electron to drift from the cathode to the anode following a 'trigger'.  A trigger can be caused by any event that would be of interest, such as the arrival of the neutrino beam, the activation of the scintillation detection system above a certain threshold, or a combination of signals from the external CRT system.  One digitization of the detector, comprised of the images of each plane for the same time window as well as all auxiliary subsystems, is referred to as an ``event''. For a typical LArTPC neutrino detector, the maximum drift time is 1-3 ms.

\begin{figure}
    \centering
    \includegraphics[width=0.5\textwidth]{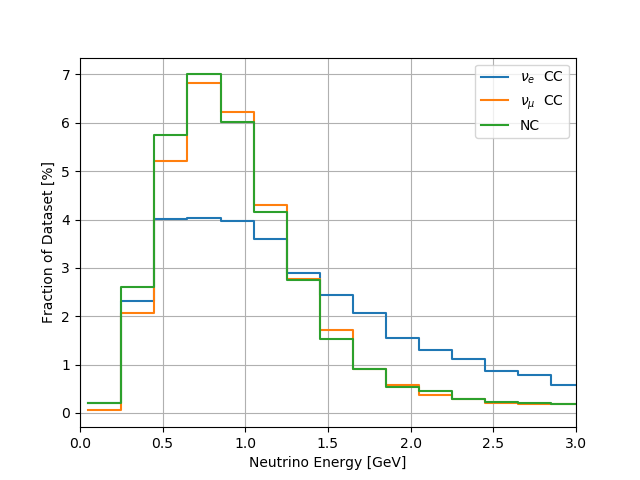}
    \caption{Neutrino energy of interactions produced for this analysis.  Most neutral current events are produced by muon-type neutrinos, and so the $\nu_\mu$ CC and Neutral Current energy spectra are similar.  The relative populations here are for the dataset used in this paper, while in the neutrino beam the muon neutrino interactions are far more frequent than electron neutrino.}
    \label{fig:sbnd_flux}
\end{figure}

The Booster Neutrino Beam delivers neutrinos to SBND up to 5 times per second, with a neutrino arrival window at the detector that is small (microseconds) compared to the TPC drift time (milliseconds).  The histogram in Figure~\ref{fig:sbnd_flux} shows the energy of interacting neutrinos simulated in SBND (more details on the simulation are in Section~\ref{sec:dataset}).  The neutrino energies range from tens of MeV to several GeV.    
When a neutrino interacts with an argon nucleus, it produces an outgoing lepton. For charged current (CC) interactions the outgoing lepton is an electron or muon for an incident electron neutrino or muon neutrino, respectively. For neutral current (NC) interactions the final state lepton is a neutrino, which exits the detector undetected.
Both kinds of interactions could also produce other particles such as pions, protons, and neutrons.  In liquid argon, at energies relevant to this work (see Figure~\ref{fig:sbnd_flux}), these particles can travel up to several meters (for energetic muons) or as little as several millimeters (for low energy protons).  

% In general, if a neutrino interacts far enough from the edges of the detector volume, SBND will observe most or all of the produced particles in the detector volume.  Typically, the only missing particles are those that exit out the side of the detector, or electrically neutral particles (such as neutrons) that can escape the liquid argon without interacting.

% Particles produced by the neutrino interactions travel through the argon at very high velocities and so the time needed to produce tracks is negligible compared to the time needed for the electron drift.  
During the few millisecond drift time of the ionization electrons, multiple incident cosmic rays will also traverse the TPC.  
Therefore, a typical event captured in coincidence with the neutrino beam has many cosmic particles visualized in the data, as seen in Figure~\ref{fig:raw_data}.
%As seen in Figure~\ref{fig:raw_data}, the number of cosmics is typically on the order of $\sim$10 events.

%The LArTPCs of the SBN program (SBND, MicroBooNE, ICARUS), which are surface detectors, 
As discussed, the scintillation light and CRT auxiliary detectors are useful for rejection of cosmic particles on a whole-image basis, but they do not have granularity to directly remove cosmic-ray induced pixels from TPC data.  For example, the photon detectors typically have spatial resolution on the order of tens of cm, while TPC data has a resolution on the scale of millimeters.  However, the temporal resolution is significantly better than the TPC. Using this timing information, which can resolve scintillation flashes coincident with the neutrino beam arrival, these detectors can easily reject non-neutrino events that have no scintillation at the right time (the neutrino-beam arrival).  

While some cosmic-only events can be rejected with light-only information, for example by requiring a flash of light coincident with the neutrino arrival from the beam, this condition is insufficient to reject every cosmic-only event.  In some cases, a neutrino can interact inside of the cryostat but external to the TPC, which is sufficient to cause a detectable flash of light in coincidence with the neutrino arrival.  However, no neutrino-induced depositions will be visible in the TPC data, even though all of the standard trigger conditions will have been met.

In another case, since each cosmic interaction also produces scintillation light in the TPC, it is possible for a cosmic particle to produce a flash of light in coincidence with the neutrino beam arrival, even if no neutrino interacts in that event.  In this case, the external cosmic ray tagger can identify the cosmic interaction in time with the beam, but these detectors have imperfect coverage and will not distinguish all in-going cosmic muons from outgoing neutrino-produced muons.

Both of these mechanisms cause an event trigger based on a flash of light during the neutrino-arrival window without any neutrino-induced activity in the TPC.  And even in events that have a neutrino interaction, the light collection and cosmic ray tagging subsystems cannot identify the neutrino interaction in the TPC data by themselves.  
Pattern recognition algorithms applied to TPC data are needed to discern cosmic-induced from neutrino-induced activity.  Traditional approaches convert TPC wire data into ``hits'' (regions of charge above noise threshold) and use geometric relationships to group hits into higher order 2D and 3D multi-hit objects within the TPC images.  These objects are treated as particles in the detector and can be further grouped with other associated objects before they are classified as being of cosmic or neutrino origin.  

In this work, we take a fundamentally different approach from traditional pattern recognition in LArTPCs by tagging the raw TPC data as cosmic-induced or neutrino-induced on a pixel-by-pixel basis.  This tagging, applied early in the analysis of TPC data, can then seed a variety of downstream analysis approaches and provide a significant boost to their performance.

% Therefore, it is desirable to have pattern recognition algorithms that are able to discern cosmic from neutrino induced activity on a pixel-by-pixel basis.  In an offline analysis, pixel-by-pixel segmentation can augment external systems to improve the rejection of cosmic-ray muon backgrounds.

%In this paper, we present novel techniques for the tagging of cosmic-induced, neutrino-induced, and background pixels using deep learning and image processing techinques.  
% The work presented here relies exclusively on pattern recognition with TPC images, and not on light collection or cosmic ray tagger systems, and is therefore an additional handle for analyzing events which pass neutrino filters based on auxiliary subsystems.

\subsection {LArTPC Imaging Data}

The individual readout ``unit'' of a LArTPC is the signal along each wire as a function of elapsed time since the trigger or event start. We form 2D images (as seen in Figure~\ref{fig:raw_data}) from the 1D wire signals as follows.
Each column of vertical pixels of the 2D image is two individual wires, one from each TPC, with the 1D signals joined at the cathode in the vertical center of the image.  Since the two TPCs drift electrons in opposite directions, away from the central cathode, the 1D signal in the top TPC is inverted compared to the bottom (here, `top' and `bottom' refer to the positions in Figure~\ref{fig:raw_data}).
The signals on each wire are juxtaposed and ordered by increasing wire location, and in this way the collection of 1D readout signals forms a high resolution 2D image.

Each constructed image is effectively a compression of 3D charge locations into a plane that runs perpendicular to every wire in the plane.
For the collection plane, with vertically oriented wires, this amounts to a top-down view of the 3D data, where the vertical information is lost in the projection.
The other two planes give a different projection, $\pm 60$ degrees from vertical, which has the effect of moving the X positions of each charge deposition, while maintaining the Y position, as compared to the vertical projection. Figure~\ref{fig:raw_data} shows the 3 wire views from the same 3D interaction in SBND.

% Of importance to this paper is the fact that the three images collected are all distinct 2D projections of the same 3D objects into different planes.  
% The three images share a common dimension in the vertical, the direction of electron drift, meaning that common objects will appear at the same height in the images.
The 3D position of a point of charge uniquely determines its location in all three images, and therefore the 3D locations of charge depositions are exactly determined from the 2D images for point-like charge.  In practice this inversion task is combinatorically hard with extended objects (and occasionally ambiguous in certain pathological topologies), but some algorithms have made excellent progress \cite{wirecell}.

\section{Related Work}
\label{sec:related}
The task of pixel level segmentation has been explored in depth in computer science journals \cite{FCN,UNet}, as well as in neutrino physics \cite{uboone_segmentation}.  In \cite{UNet}, shortcut connections are introduced to a fully convolutional segmentation network for biological images.  The network we present in this paper is similar to the `UNet' architecture in that it has shortcut connections between down-sampled and up-sampled layers of similar resolution.  More details of the building blocks and architecture are given in Section~\ref{sec:network}.

In \cite{uboone_segmentation}, a modified version of UNet, using residual convolution layers, was deployed to perform pixel-level segmentation of particles based on particle topology; electrons and photons exhibited a broader, ``fuzzy'' topology when compared to ``track''-like particles (protons, muons, pions) which typically are seen as thin, line-like objects.\footnote{The ``fuzzy''ness of electromagnetic particles is due to the electromagnetic cascade or shower of particles initiated by an electron or photon with enough energy to produce more particles.}  The network was trained for 512x512 square images of data from the MicroBooNE detector, and the result was a successful first application of UNet style segmentation techniques to  LArTPC neutrino data.  Following \cite{uboone_segmentation}, the network described in this paper also applies a series of residual blocks instead of pure convolutions at each image resolution, hence is referred to as `UResNet'.

Additionally, in \cite{domin2019scalable}, the authors introduce a spatially sparse, UResNet style architecture for particle-wise segmentation labels in both a 2D and 3D LArTPC-like dataset.  Their result is based purely on \texttt{GEANT4}~\cite{geant} information, meaning that the images did not include the simulation of electronic effects, nor drift-induced effects such as diffusion or absorption of electrons.
Nevertheless, this is a novel technique that has broad applicability in neutrino physics.  The results presented here use a dense convolutional network, however it is notable that a sparse implementation of the results presented here could deliver gains in performance and computational efficiency.

In MicroBooNE analyses, classical reconstruction techniques are used to reject cosmic ray particles on a particle-by-particle basis, after particles have been ``reconstructed'' into distinct entities with traditional pattern recognition analyses.  For example, in an analysis of charged current muon neutrino \cite{microboone_numu} there is still a background of approximately 35\% cosmic or cosmic-contaminated interactions at 50\% signal selection efficiency.  
% More complicated topologies, such as electron neutrino interactions and neutral current interactions, are likely to suffer even higher contamination rates at lower selection efficiencies.  
The results presented here have been developed with the SBND TPC and geometry in mind, but should apply well to the MircoBooNE or ICARUS geometries, also along the Booster Neutrino Beam and part of the SBN Program.  In general,, the techniques presented here are intended to augment analyses such as \cite{microboone_numu} to gain better background rejection and better signal efficiency.
\section{Dataset}
\label{sec:dataset}
% The dataset for this application was generated via the \texttt{larsoft} \cite{larsoft} package and in particular the SBND geometry and detector simulation in 2018.  It is known that the electronics simulation for SBND are not finalized, and the geometry definition was also not finalized at the creation of these files.  While this detracts from the immediate applicability of the dataset to real data, the techniques to create the dataset and train the network are directly applicable when the final simulation will be ready.  Minor changes to the geometry and electronics response are unlikely to lead to large changes in performance.
The dataset for this application was generated via the \texttt{larsoft} simulation toolkit for LArTPCs~\cite{larsoft} utilizing a SBND geometry description and electronics simulation, as of 2018.  It was known that the geometry description and electronics simulation for SBND were not finalized at that time, but minor changes to the geometry and electronics response are unlikely to lead to significant changes in the performance we report here.

The drift direction in each plane is 
digitized at a higher spatial resolution than the wire spacing.  For this dataset, the images are downsampled along the drift direction by a factor of 4 to make vertical and horizontal distances have the same scale. 
To better suit downsampling and upsampling operations, the images are centered horizontally into images with a width of 2048 pixels, with each pixel representing one wire. The drift direction is 1260 pixels.
Pixels on the right and left, beyond the original image, are set to 0 in both label and input images.  The cathode is visible in these images as a green horizontal space in the middle of each image.

% original width was 0.49
\begin{figure}
\begin{centering}
    \includegraphics[width=0.6\columnwidth]{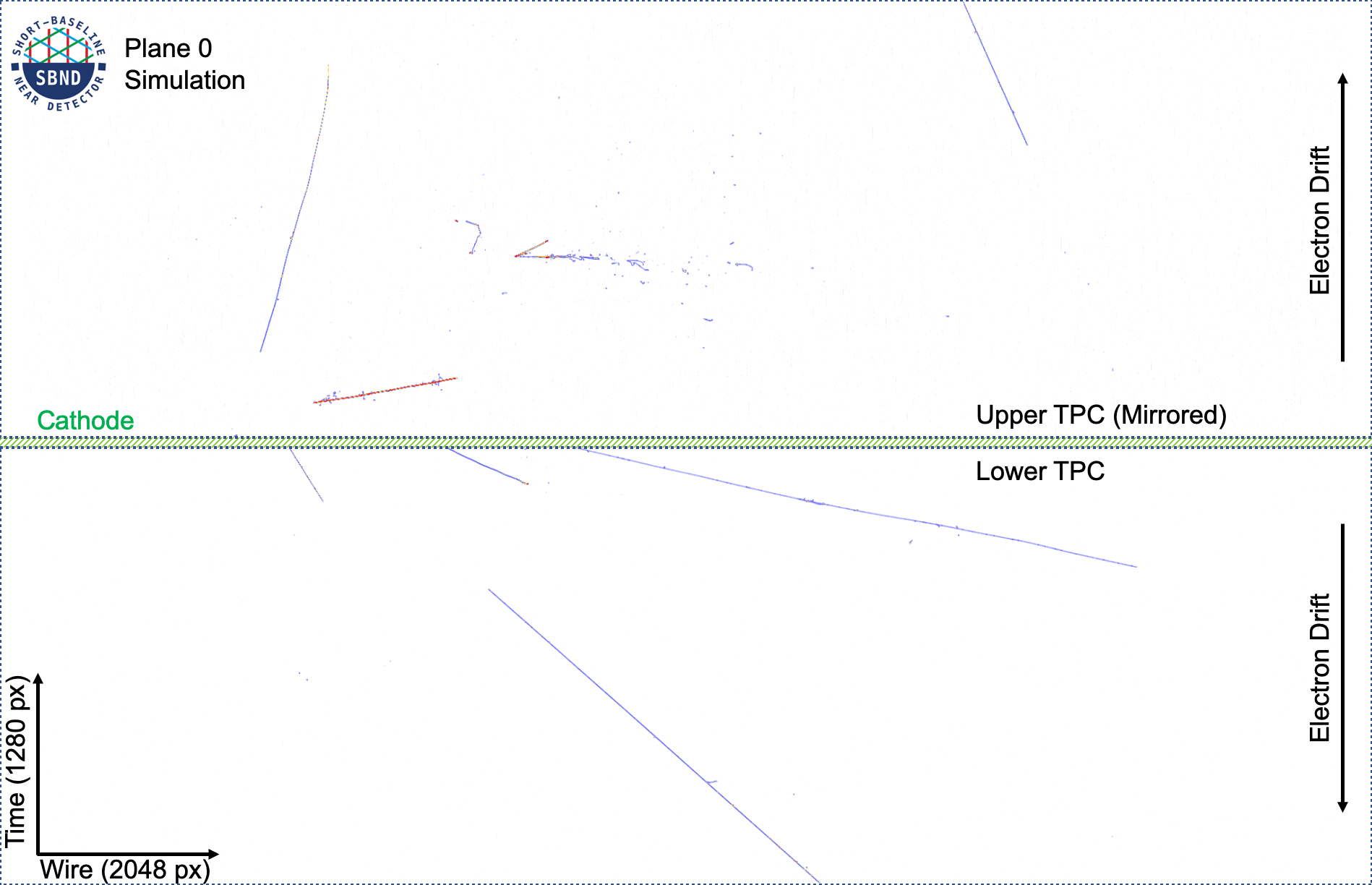}
    \includegraphics[width=0.6\columnwidth]{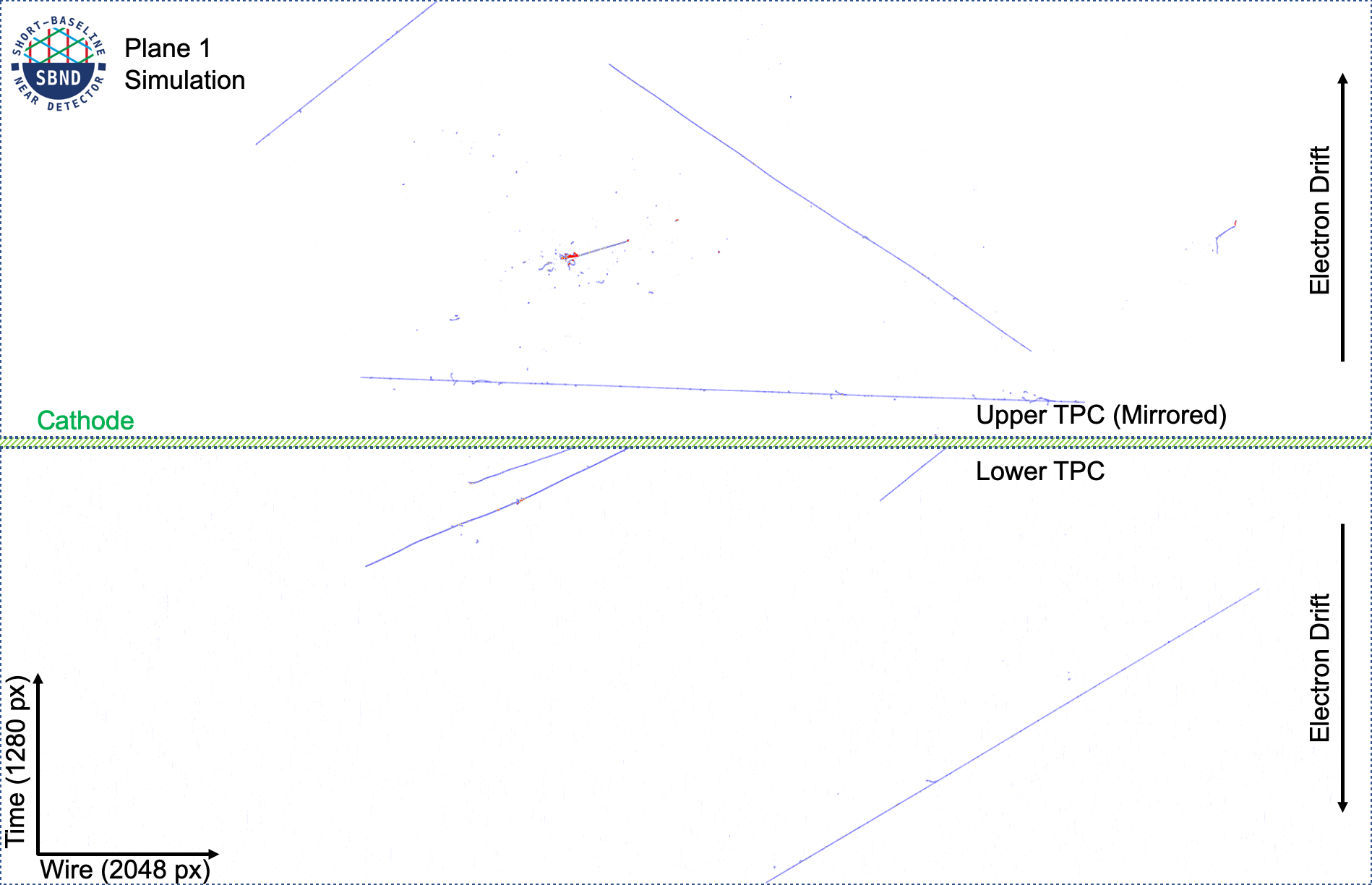}
    \includegraphics[width=0.6\columnwidth]{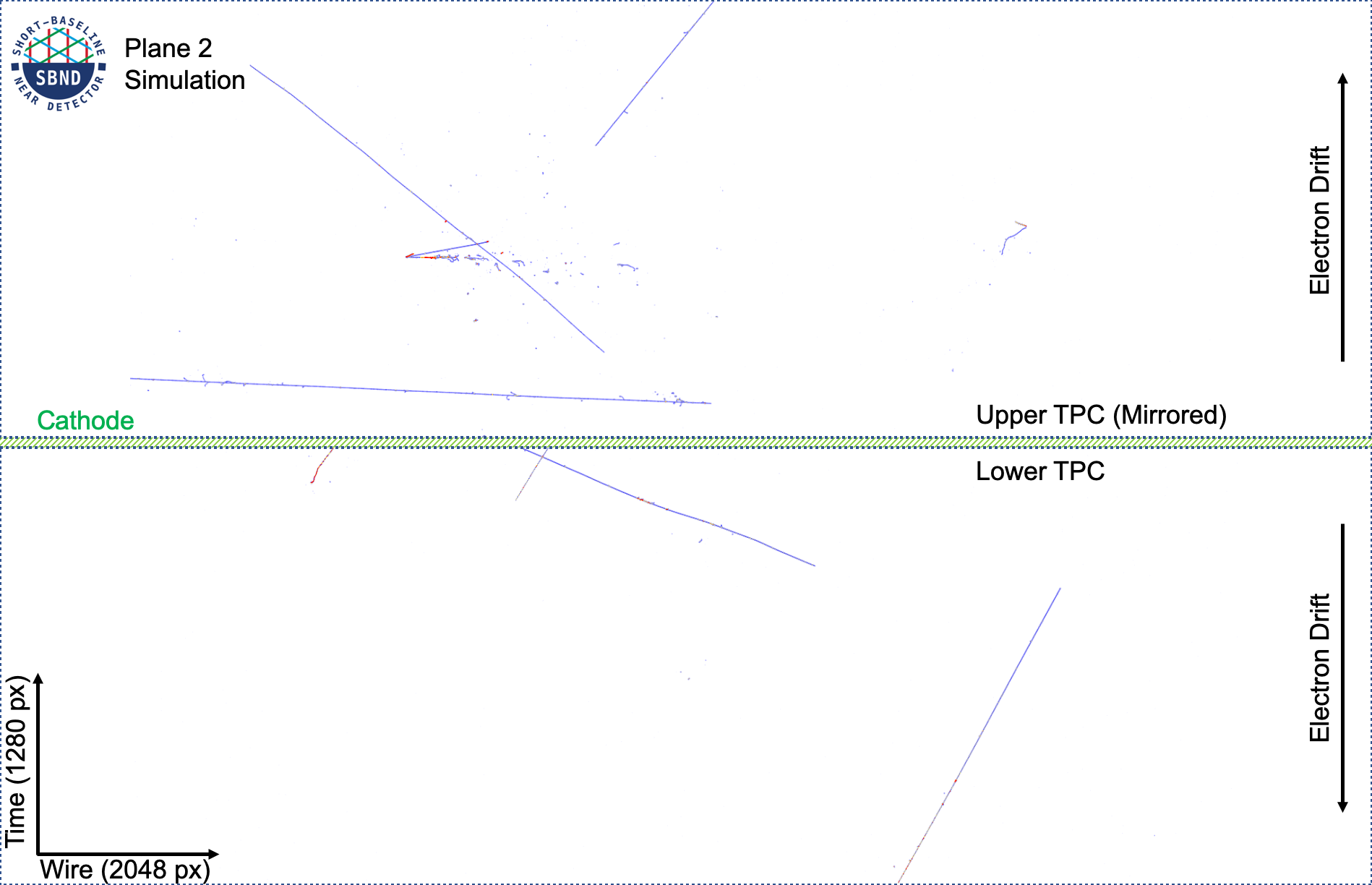}
    \caption{The raw data for one image in the dataset at full resolution. Charge observed is colored with blue for smaller charge depositions and red for larger charge depositions.  There is an electron neutrino charged current interaction in the Upper TPC.}
    \label{fig:raw_data}
\end{centering}
\end{figure}

Because the images to segment are so large, this work is demonstrating these results on a downsampled version of the images, where each image is at 50\% resolution (640 pixels tall, 1024 pixels wide).

Each interaction in the dataset used here has neutrino interactions simulated with the GENIE software package \cite{genie} (v2.12.8c), and cosmic backgrounds simulated using CORSIKA \cite{corsika} (v1.7i).  The BNB neutrino flux is used to sample neutrinos at the proper energies, however the relative populations of three distinct categories of events ($\nu_\mu$ CC, $\nu_e $ CC, and NC) are balanced in the training set (see Figure~\ref{fig:sbnd_flux}).

\begin{figure}
\begin{centering}
    \includegraphics[width=0.6\columnwidth]{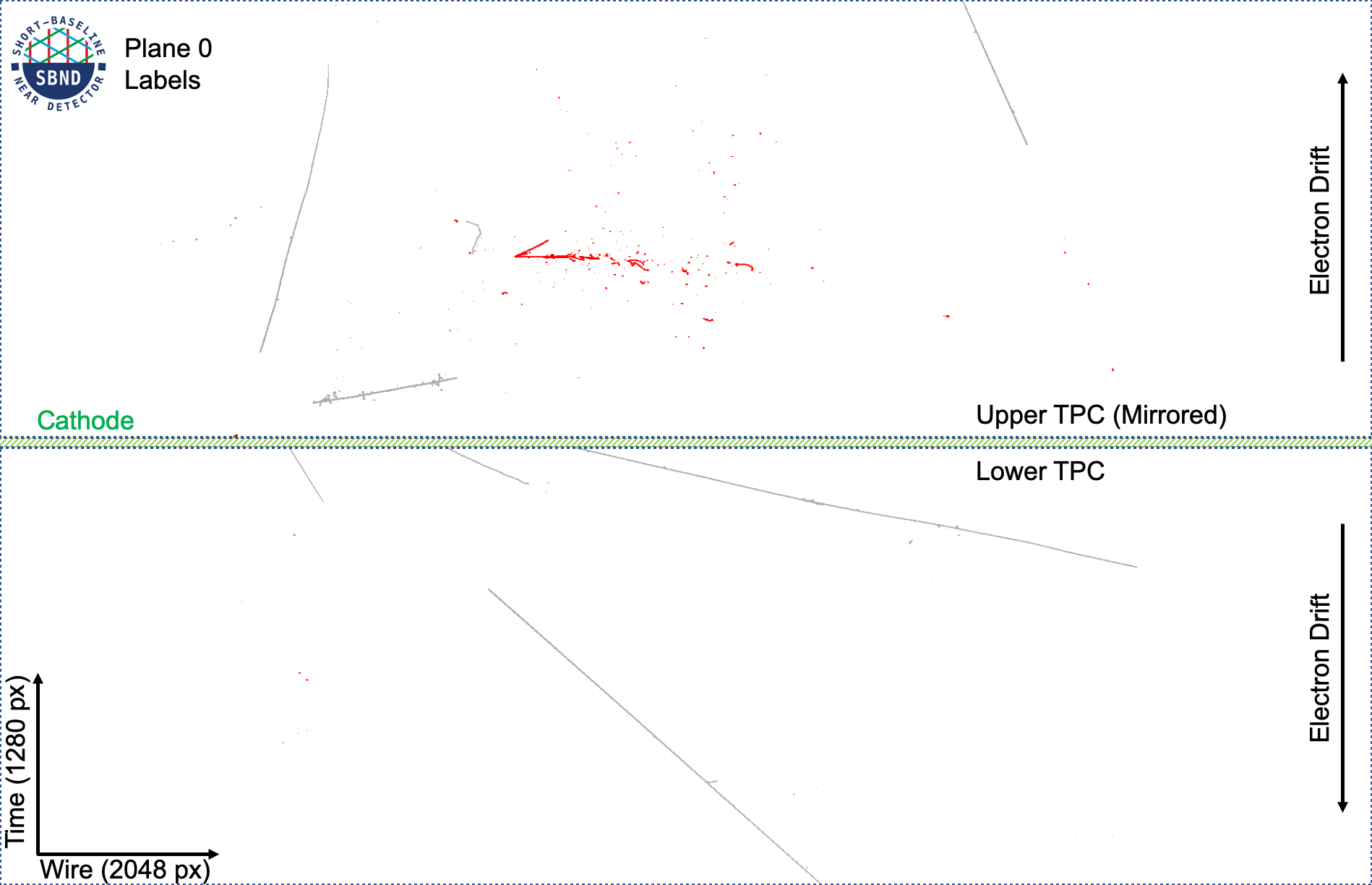}
    \includegraphics[width=0.6\columnwidth]{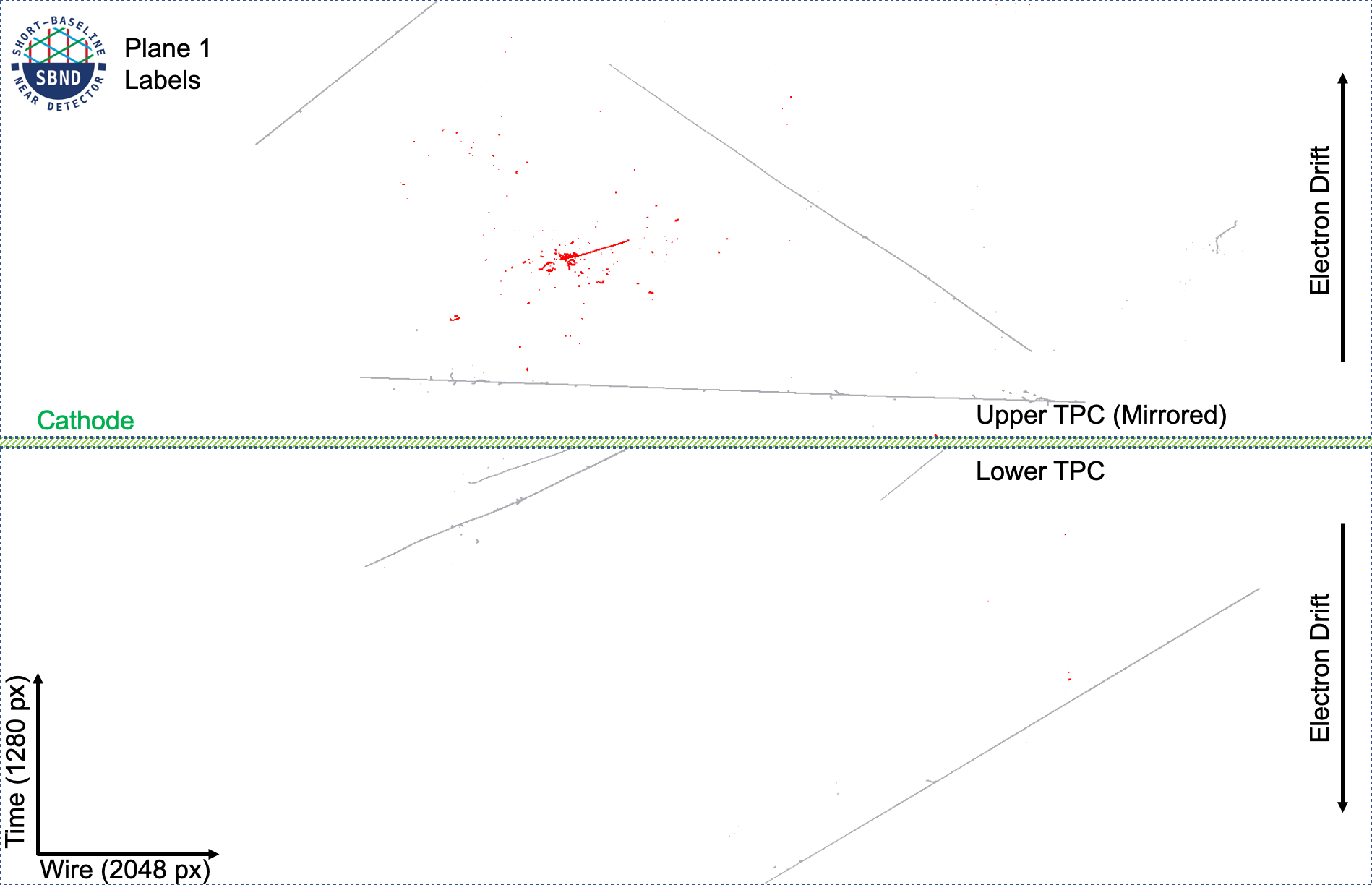}
    \includegraphics[width=0.6\columnwidth]{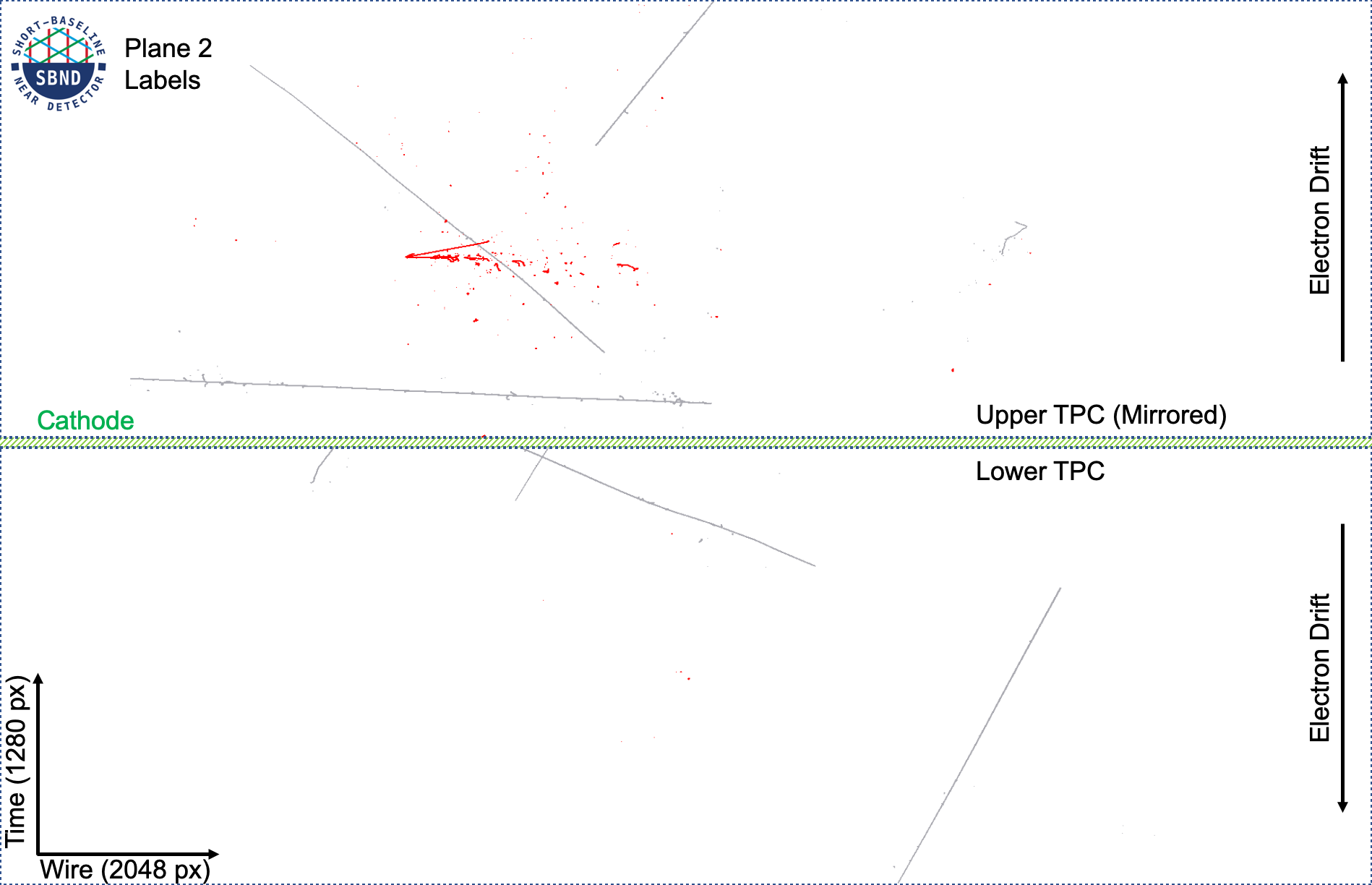}
    \caption{The labels for the images in Figure~\ref{fig:raw_data} in the dataset at full resolution.  White pixels are background, gray pixels are associated with cosmic particles, and red pixels are associated with a neutrino interaction.  Plane 2 shows a case of overlap between cosmic and neutrino pixels.}
    \label{fig:labeled_data}
\end{centering}
\end{figure}

The label images are created using truth level information from \texttt{GEANT4}~\cite{geant} (v4.10.3.p01b), where each deposition on a wire is tracked from the particle that created it.  Each particle, in turn, is tracked to its parent particle up to the primary particles.  All depositions that come from a particle (or its ancestor) that originated with GENIE are labeled as neutrino induced, and all depositions that originated from a CORSIKA particle are labeled as cosmics.  In the event of an overlap, as is common, the neutrino label takes precedence.  Approximately 50\% of all events have an overlap in at least one plane.  The label images for the event in Figure~\ref{fig:raw_data} can be seen in Figure~\ref{fig:labeled_data}.

\section{Network Architectures and Implementations}
\label{sec:network}

For this work, we present a novel modification of the UResNet architecture for cosmic and neutrino segmentation that aims to meet several criteria:

\begin{itemize}
    \item Discriminate cosmic pixels from neutrino pixels with high granularity.
    \item Segment entire events across all planes simultaneously and efficiently.
    \item Incorporate multi-plane geometrical information.
\end{itemize}

\begin{figure}
    \includegraphics[width=\columnwidth]{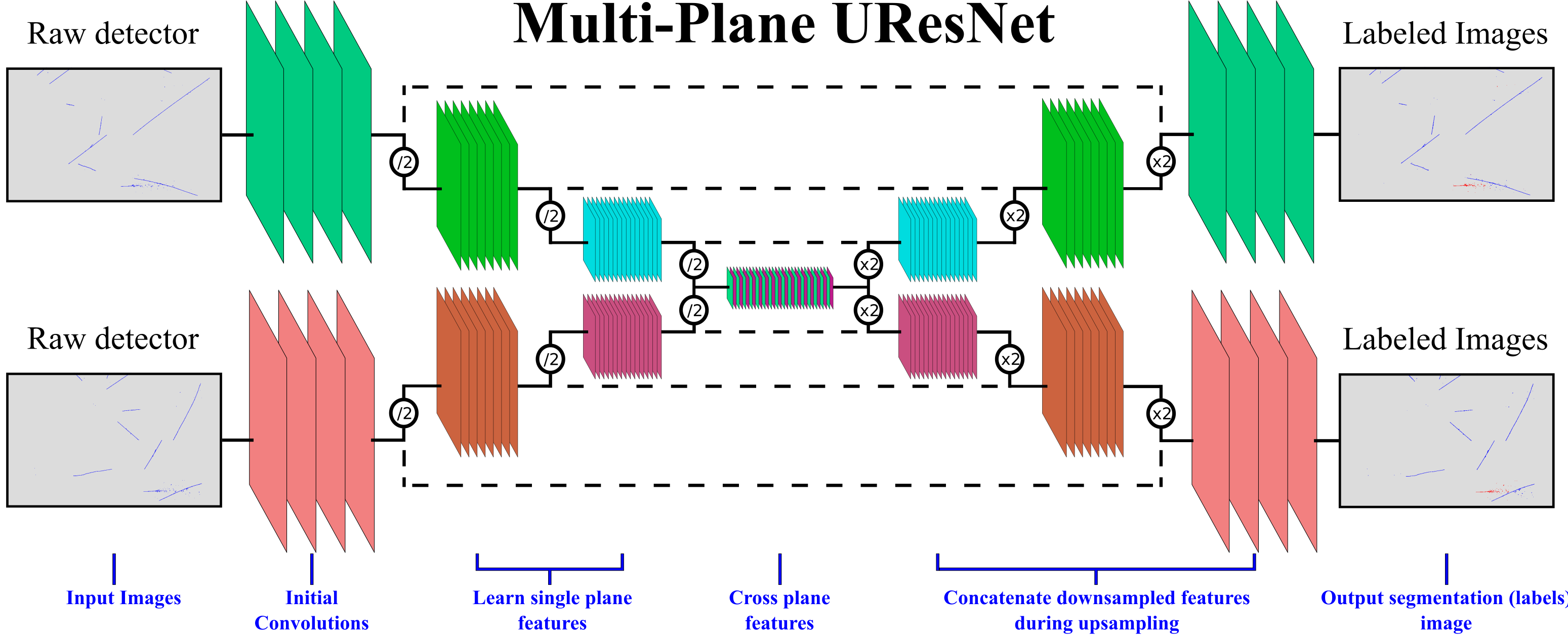}
    \caption{A representation of the multi-plane UResNet architecture.  Only two of the three planes are shown in this image for clarity.}
    \label{fig:multiplane-uresnet}
\end{figure}

To this end, we present a multi-plane, UResNet style architecture as depicted in Figure~\ref{fig:multiplane-uresnet}.  The input to the network is entire images for each of the 3 planes, each of which is fed through a segmentation network in the shape of a UResNet.  Unique to this work, at the deepest convolutional layer, the per-plane filters are concatenated together into one set of convolutional filters and proceed through convolutions together, in order to learn cross-plane geometrical features.  Without this connection at the deepest layer, this network is exactly a ``standard'' UResNet architecture applied to each plane independently.  We see in our experimental results below that without this connection layer, the network does not perform as well.  After this, the filters are split and up-sampled independently again.

Because each plane has similar properties at a low level (i.e., particles look similar in each plane, even if the geometric projection is different), convolutional weights are shared across all three planes for up-sampling and down-sampling of the network.

The implementation of the network is available in both TensorFlow~\cite{tensorflow2015-whitepaper} and PyTorch \cite{pytorch} on GitHub\footnote{\href{https://github.com/coreyjadams/CosmicTagger}{https://github.com/coreyjadams/CosmicTagger}}.  The basic building blocks of this network are residual convolutional layers \cite{resnet}.  In a residual layer, the input tensor is processed with convolutions, non-linear activations, and (potentially) normalization layers before being summed with the input of residual layer: $R(x) = x + C(x)$, where $R$ is the residual function and $C$ represents the convolution layers.  In this work, we use Batch Normalization \cite{batch_norm} as a normalization layer, and LeakyReLU~\cite{leaky_relu} as a non-linear activation. While there are many configuration parameters, the baseline model has 6 levels of depth and the following properties:

\begin{itemize}
    \item The network operates on each plane independently except at the very deepest layer.
    \item The first layer of the network is a 7x7 convolutional filter that outputs a parametrizable number of filters - the reference models use 16.
    \item Each subsequent layer in the down-sampling pass takes the previous output and applies two residual blocks, described below, followed by a max pooling to reduce the spatial size.  After the max pooling, a bottleneck 1x1 convolution increases the number of filters by a factor of 2.
    \item After the 5th down-sampling pass, the spatial size of the images is (10,16) with 512 filters in each plane.  The images from each plane are concatenated together, and a bottleneck convolution is applied across the concatenated tensor to reduce the number of filters to 256.  Then, 5 residual blocks of size 5x5 are applied, followed by a 1x1 layer to increase the number of filters back to 1536.  The filters are split into three tensors again.
    \item After the deepest layer, each up-sampling layer takes the output of the corresponding downward pass, adds it to the output of the previous up-sampling layer, and performs two residual blocks with 3x3 convolutions.  This pattern of up-sampling/addition/convolutions continues until original resolution is reached.
    \item Once the original resolution has been restored, a single 1x1 convolution is applied to output 3 filters for each image, where the 3 filters correspond to the 3 background classes.
\end{itemize}

The details of each layer are summarized in Table~\ref{tab:network}.  The residual blocks used in the network mirror those in \cite{resnet}, and are the following sequence of operations: convolution, Batch Normalization, LeakyReLU, convolution, Batch Normalization, sum with input, LeakyReLU.

\begin{table}
\begin{centering}
\begin{tabular}{rccccl}
\hline
\multicolumn{1}{|r|}{Layer}      & \multicolumn{1}{c|}{X}   & \multicolumn{1}{c|}{Y}    & \multicolumn{1}{c|}{Filters} & \multicolumn{1}{c|}{Parameters} & \multicolumn{1}{l|}{Operations}                                                 \\ \hline
\multicolumn{1}{|r|}{Initial}    & \multicolumn{1}{c|}{640} & \multicolumn{1}{c|}{1024} & \multicolumn{1}{c|}{1}       & \multicolumn{1}{c|}{416}        & \multicolumn{1}{l|}{conv7x7, BN, LeakyReLU}                                          \\ \hline
\multicolumn{1}{|r|}{Down 0}     & \multicolumn{1}{c|}{640} & \multicolumn{1}{c|}{1024} & \multicolumn{1}{c|}{8}       & \multicolumn{1}{c|}{2576}       & \multicolumn{1}{l|}{Res3x3, Res3x3, MaxPool, Bottleneck 8 to 16}                \\ \hline
\multicolumn{1}{|r|}{Down 1}     & \multicolumn{1}{c|}{320} & \multicolumn{1}{c|}{512}  & \multicolumn{1}{c|}{16}      & \multicolumn{1}{c|}{10016}      & \multicolumn{1}{l|}{Res3x3, Res3x3, MaxPool, Bottleneck 16 to 32}               \\ \hline
\multicolumn{1}{|r|}{Down 2}     & \multicolumn{1}{c|}{160} & \multicolumn{1}{c|}{256}  & \multicolumn{1}{c|}{32}      & \multicolumn{1}{c|}{39488}      & \multicolumn{1}{l|}{Res3x3, Res3x3, MaxPool, Bottleneck 32 to 64}               \\ \hline
\multicolumn{1}{|r|}{Down 3}     & \multicolumn{1}{c|}{80}  & \multicolumn{1}{c|}{128}  & \multicolumn{1}{c|}{64}      & \multicolumn{1}{c|}{156800}     & \multicolumn{1}{l|}{Res3x3, Res3x3, MaxPool, Bottleneck 64 to 128}              \\ \hline
\multicolumn{1}{|r|}{Down 4}     & \multicolumn{1}{c|}{40}  & \multicolumn{1}{c|}{64}   & \multicolumn{1}{c|}{128}     & \multicolumn{1}{c|}{624896}     & \multicolumn{1}{l|}{Res3x3, Res3x3, MaxPool, Bottleneck 128 to 256}             \\ \hline
\multicolumn{1}{|r|}{Down 5}     & \multicolumn{1}{c|}{20}  & \multicolumn{1}{c|}{32}   & \multicolumn{1}{c|}{256}     & \multicolumn{1}{c|}{2494976}    & \multicolumn{1}{l|}{Res3x3, Res3x3, MaxPool, Bottleneck 256 to 512}             \\ \hline
\multicolumn{1}{|r|}{Bottleneck} & \multicolumn{1}{c|}{10}  & \multicolumn{1}{c|}{16}   & \multicolumn{1}{c|}{1536}    & \multicolumn{1}{c|}{393984}     & \multicolumn{1}{l|}{Concat across planes, bottleneck 1536 to 256}               \\ \hline
\multicolumn{1}{|r|}{Deepest}    & \multicolumn{1}{c|}{10}  & \multicolumn{1}{c|}{16}   & \multicolumn{1}{c|}{256}     & \multicolumn{1}{c|}{16391680}   & \multicolumn{1}{l|}{Res5x5, 5 layers}                                           \\ \hline
\multicolumn{1}{|r|}{Bottleneck} & \multicolumn{1}{c|}{10}  & \multicolumn{1}{c|}{16}   & \multicolumn{1}{c|}{1536}    & \multicolumn{1}{c|}{397824}     & \multicolumn{1}{l|}{bottleneck 256 to 1536, split into 3 planes}                \\ \hline
\multicolumn{1}{|r|}{Up 5}       & \multicolumn{1}{c|}{20}  & \multicolumn{1}{c|}{32}   & \multicolumn{1}{c|}{256}     & \multicolumn{1}{c|}{2494208}    & \multicolumn{1}{l|}{Interp., Sum w/ Down 5, Bottleneck, Res3x3, Res3x3} \\ \hline
\multicolumn{1}{|r|}{Up 4}       & \multicolumn{1}{c|}{40}  & \multicolumn{1}{c|}{64}   & \multicolumn{1}{c|}{128}     & \multicolumn{1}{c|}{624512}     & \multicolumn{1}{l|}{Interp., Sum w/ Down 4, Bottleneck, Res3x3, Res3x3} \\ \hline
\multicolumn{1}{|r|}{Up 3}       & \multicolumn{1}{c|}{80}  & \multicolumn{1}{c|}{128}  & \multicolumn{1}{c|}{64}      & \multicolumn{1}{c|}{156608}     & \multicolumn{1}{l|}{Interp., Sum w/ Down 3, Bottleneck, Res3x3, Res3x3} \\ \hline
\multicolumn{1}{|r|}{Up 2}       & \multicolumn{1}{c|}{160} & \multicolumn{1}{c|}{256}  & \multicolumn{1}{c|}{32}      & \multicolumn{1}{c|}{39392}      & \multicolumn{1}{l|}{Interp., Sum w/ Down 2, Bottleneck, Res3x3, Res3x3} \\ \hline
\multicolumn{1}{|r|}{Up 1}       & \multicolumn{1}{c|}{320} & \multicolumn{1}{c|}{512}  & \multicolumn{1}{c|}{16}      & \multicolumn{1}{c|}{9968}       & \multicolumn{1}{l|}{Interp., Sum w/ Down 1, Bottleneck, Res3x3, Res3x3} \\ \hline
\multicolumn{1}{|r|}{Bottleneck} & \multicolumn{1}{c|}{640} & \multicolumn{1}{c|}{1024} & \multicolumn{1}{c|}{16}      & \multicolumn{1}{c|}{2552}       & \multicolumn{1}{l|}{Bottleneck1x1 to 3 output filters.}                         \\ \hline
\multicolumn{1}{|r|}{Final}      & \multicolumn{1}{c|}{640} & \multicolumn{1}{c|}{1024} & \multicolumn{1}{c|}{3}       & \multicolumn{1}{c|}{57}         & \multicolumn{1}{l|}{Final Segmentation Maps}                                    \\ \hline
\multicolumn{1}{l}{}             & \multicolumn{1}{l}{}     & \multicolumn{1}{l}{}      & \multicolumn{1}{l}{}         & \multicolumn{1}{l}{}            &
\end{tabular}
\caption{A description of the multi-plane UResNet architecture used in this work.}
\label{tab:network}
\end{centering}
\end{table}

To summarize, the network architecture used here is taking state-of-the-art segmentation techniques (`UNet'~\cite{UNet} and `UResNet'~\cite{uboone_segmentation}) and enhancing them to learn correlated features across images.

\subsection{Analysis Metrics}

Because of the sparse nature of the images from a LArTPC detector, the per-pixel accuracy does not give good discriminating power to gauge network performance. Simply predicting `background' for all pixels yields a very high accuracy over 99\% - even with every `cosmic' and `neutrino' pixel mislabeled.  To mitigate this, we calculate several metrics that have proven useful for measuring the performance of a cosmic tagging network:

\begin{itemize}
\item {\bf Accuracy} is computed as the total fraction of pixels that are given the correct label by the network, where the predicted label is the highest scoring category in the softmax for that pixel.
\item {\bf Non-background Accuracy} is the same as Accuracy above, but computed only for pixels that have a non-zero label in the truth labels.  In basic terms, this metric is measuring how often the network is predicting the correct pixel on the parts of the image that matter, as background pixels can easily be identified from their lack of charge.
\item {\bf Intersection over Union} (or IoU) is calculated for the neutrino (and cosmic) pixels.  This metric uses the set of pixels that are {\em labeled} (by the simulation) as neutrino (or cosmic) and the set of pixels that are {\em predicted} (by the network) as neutrino (or cosmic).  The metric is the ratio of the number of pixels that are in both sets (intersection) divided by the number of pixels in either set (union).  
% $ IoU \equiv \frac{X \cap Y }{x \cup Y}$
In basic terms, this metric measures how often the network predicts active categories (neutrino, cosmic) on the correct pixels and {\bf only} the correct pixels.
\end{itemize}

\section{Training}
\label{sec:training}

\begin{figure}
\begin{centering}
    \includegraphics[width=0.6\columnwidth]{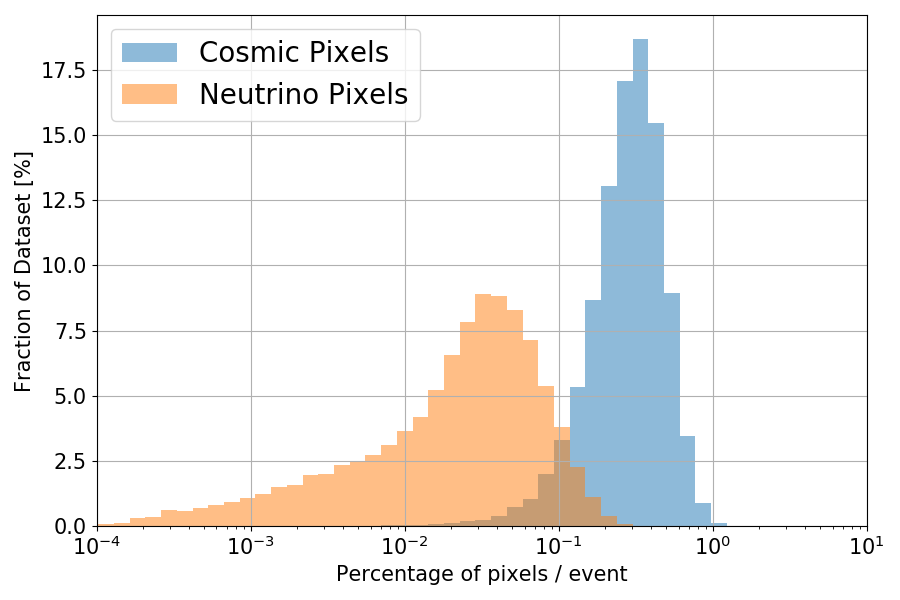}
    \caption{Distribution of pixel occupancies, by label, in this dataset.  In general, the cosmic-labeled pixels are less than 1\% of pixels and the neutrino-labeled pixels are less than 0.3\%}
    \label{fig:pixel_occupancy}
\end{centering}
\end{figure}

The network here is trained on a down-sampled version of the full-event images, so each event represents three planes of data at a height of 640 pixels and a width of 1024 pixels, for a total of 655,360 pixels per plane and 3 planes.  Though it would be ideal to train on full-resolution images, this is prohibitive computationally as the network doesn't fit into RAM on current generation hardware.

The number of active (non-zero) pixels varies from image to image. In general the number of pixels which have some activity, either from particle interactions or simulated noise, is approximately 11,000 per plane.  Of these, approximately 2300 per plane on average are from cosmic particles, and merely $\sim$250 per plane are from neutrino interactions, on average.  See Figure~\ref{fig:pixel_occupancy} for more details.

To speed up training and ensure the neutrino pixels, which are the most important scientifically, are well classified, we adopt a weight scaling technique.  The loss for each pixel is a 3 category cross entropy loss, and the traditional loss per plane would be the average over all pixels in that plane. Here, instead, we boost the loss of cosmic pixels by a factor of 1.5, and neutrino pixels by a factor of 10.  The final loss is averaged over all pixels in all three planes.  We also experimented with a loss-balancing technique where, in each image, the weight for each pixel is calculated so the product of the total weight of all pixels in each category is balanced: $weight_{background} \times N_{background} = weight_{cosmic} \times N_{cosmic} = weight_{neutrino} \times N_{neutrino}$. Experimentally, we find that more aggressive loss boosting of neutrino and cosmic pixels leads to blurred images around the cosmic and neutrino pixels, as those pixels are heavily de-weighted as background pixels.  In future studies, we plan to investigate the use of dynamic loss functions such as focal loss~\cite{focal_loss} to allow better balancing of background to significant pixels throughout training.

We report here the performance of several variations of the network, in order to examine the properties of the final accuracy and determine the best network.  We test several variations of the network.  The baseline model is as described above, trained with the mild weight balancing, using an RMSProp \cite{rmsprop} optimizer.  For variations we train the same network with the following modifications:
\begin{itemize}
    \item \textbf{Concatenated Connections} - instead of additive connections across the ``U'', we use concatenation and 1x1 convolutions.
    \item \textbf{Cross-plane Blocked} - the concat operation blocked at the deepest layer (no cross-plane information), effectively using a single-plane network 3 times simultaneously.
    \item \textbf{Batch Size $\times$ 2 } - a minibatch size of 16, instead of 8, is used.
    \item \textbf{Convolutional Upsample} - convolutional up-sampling instead of interpolation up-sampling.
    \item \textbf{Num. Filters / 2} - fewer initial filters (8 instead of 16).
    \item \textbf{No Loss Balance} - all pixels are weighted equally without regard to their label.
    \item \textbf{Larger Learning Rate} - the learning rate is set to 0.003 (10x higher).
    \item \textbf{Non Residual} - no residual connections in the down-sampling and up-sampling pass.
    \item \textbf{Adam Optimizer} - unmodified network trained with Adam Optimizer \cite{Adam}.
    \item \textbf{Full Balance} - a full loss balancing scheme where each category is weighted such that the sum across pixels of the weights for each category is 1/3.
\end{itemize}

\begin{figure}
    \centering
    \includegraphics[width=0.49\textwidth]{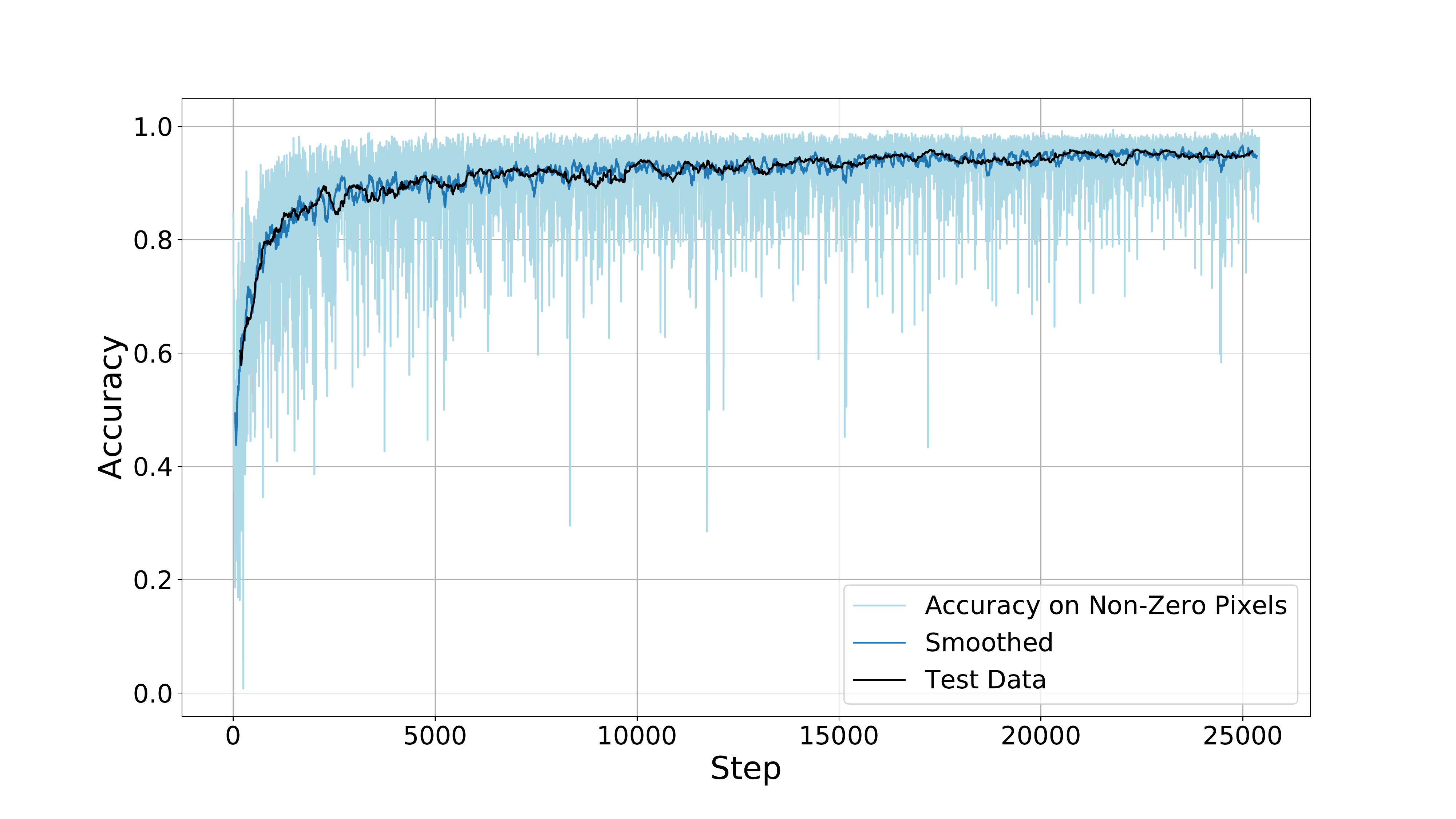}
    \includegraphics[width=0.49\textwidth]{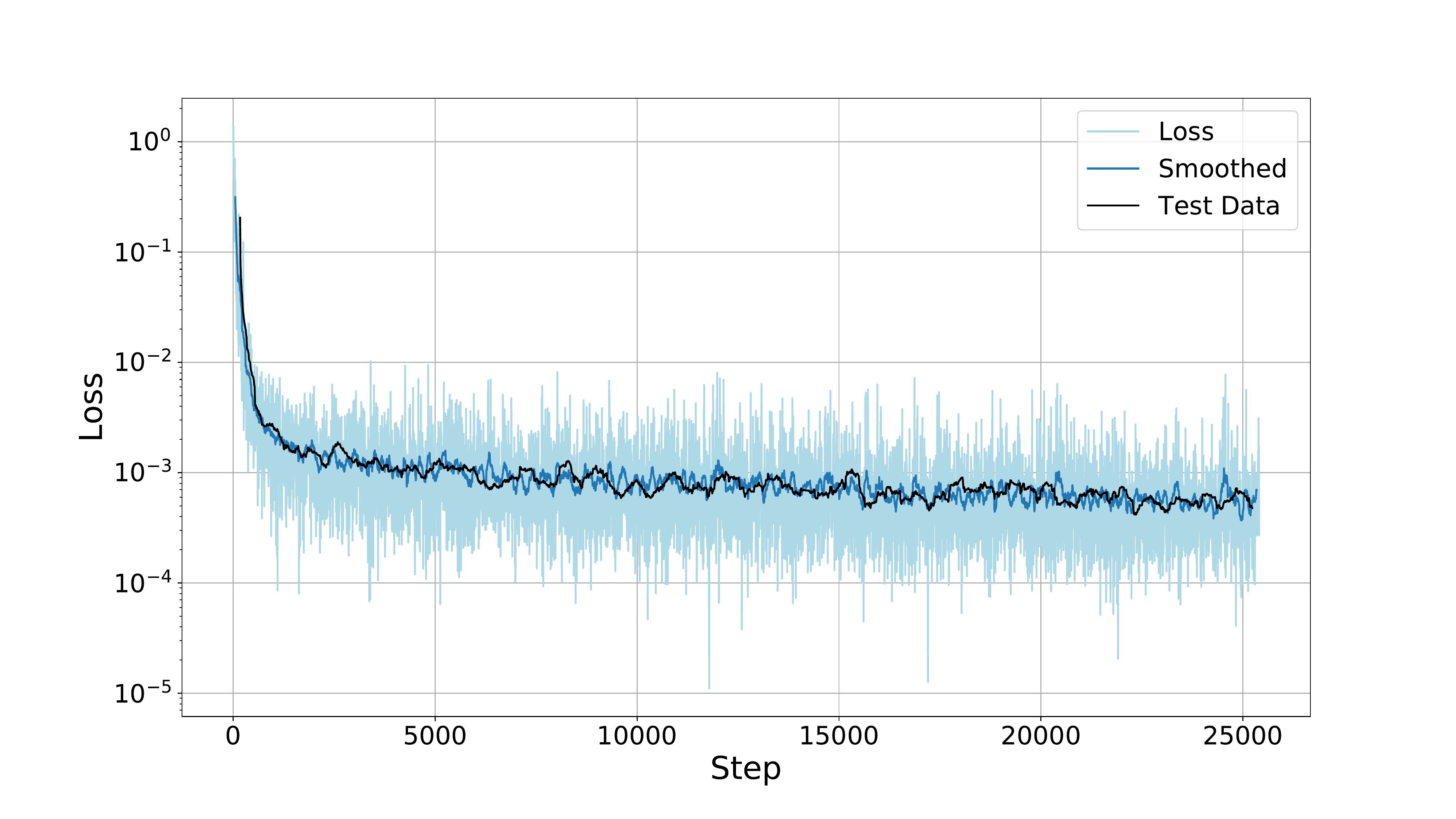}
    \includegraphics[width=0.49\textwidth]{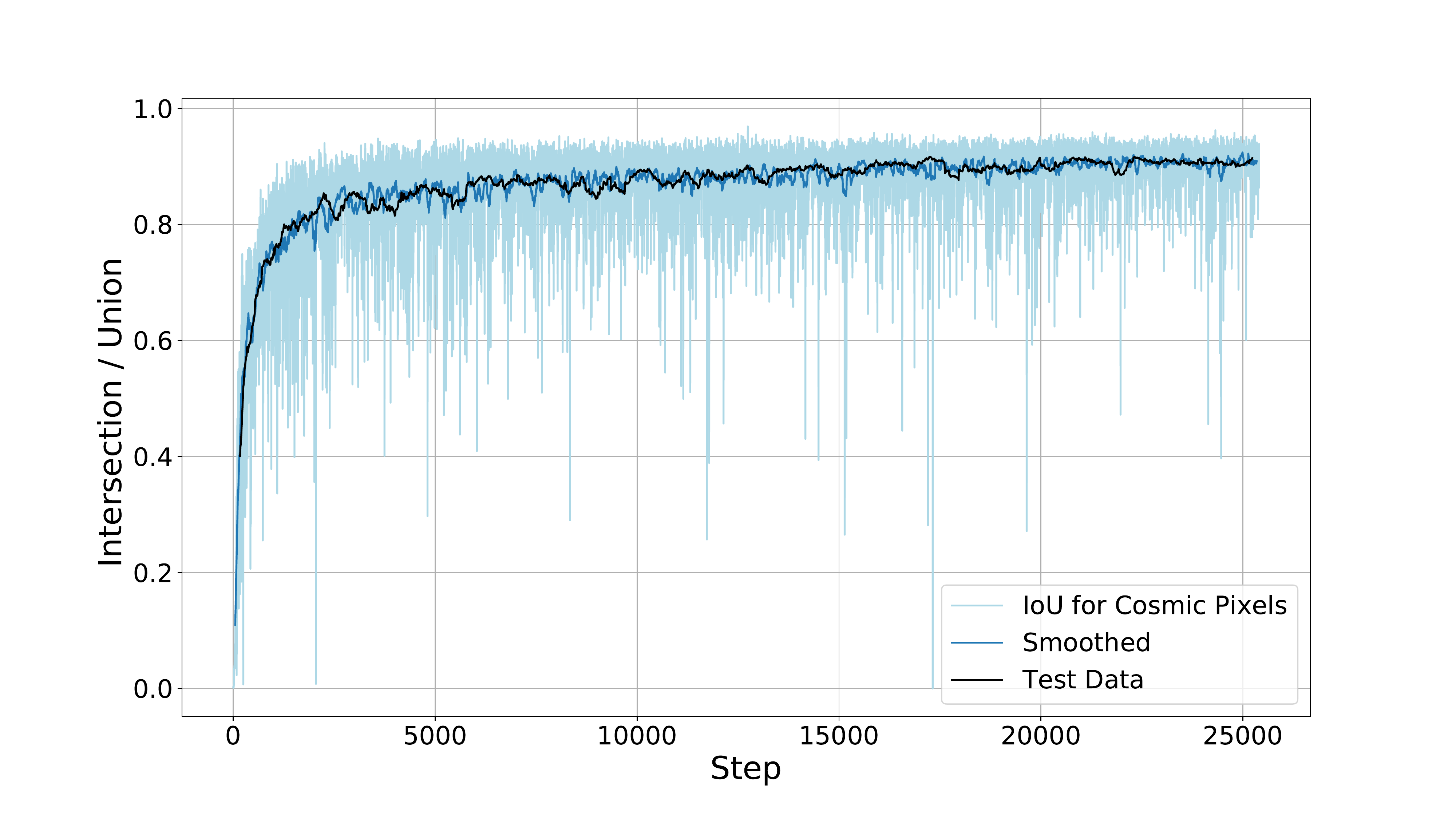}
    \includegraphics[width=0.49\textwidth]{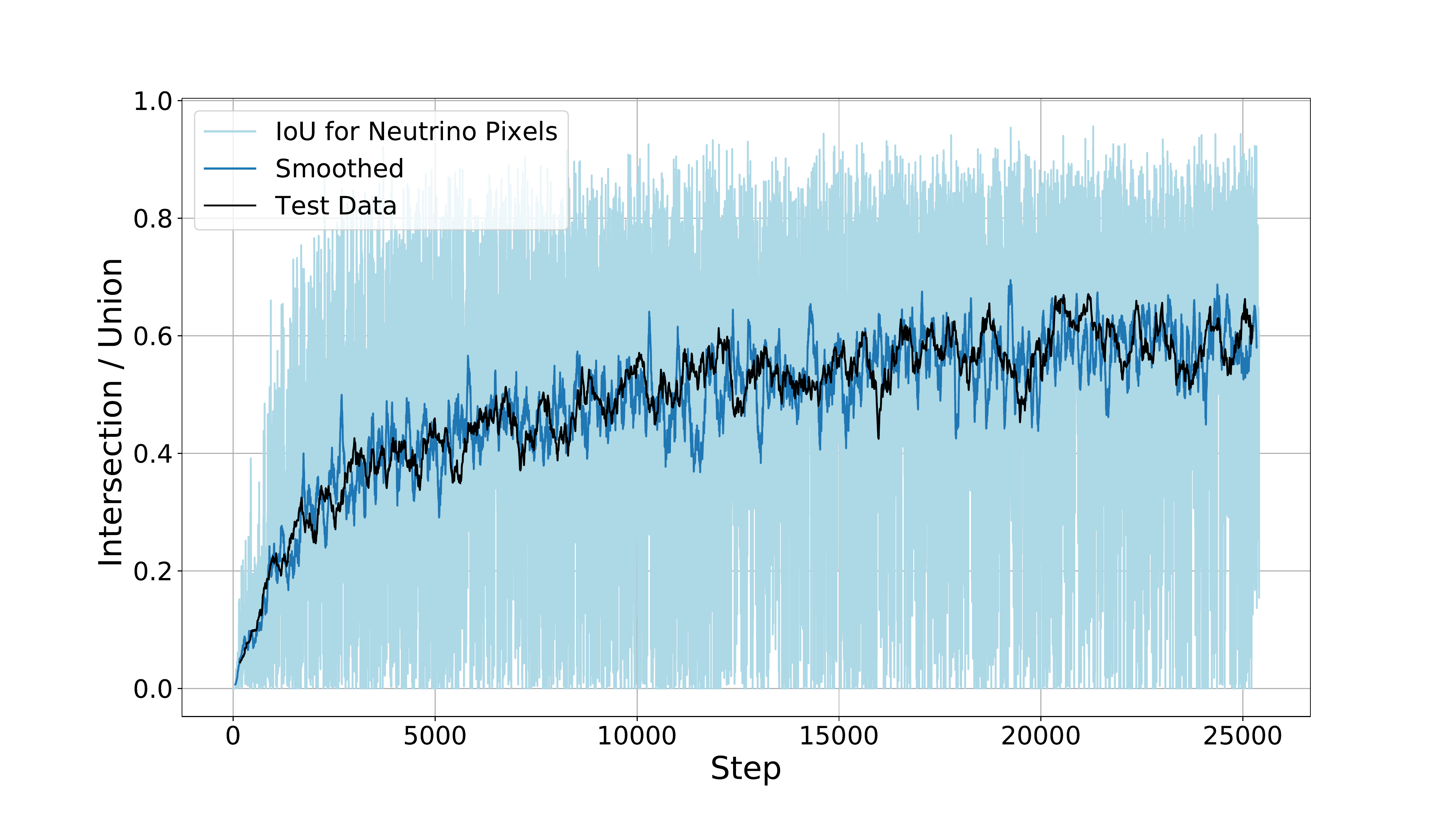}
    \caption{The training progression of the baseline model, trained for 25k iterations. The light blue curve is the training performance at each step, overlaid with a smoothed representation of the same data, and a smoothed representation of the test set.}
    \label{fig:training_metrics}
\end{figure}

All models, except one, are trained with a minibatch size of 8 ($\times$ three images, one per plane).  The learning rate is set to 0.0003, except for the network that uses a higher learning rate.  The other network is trained with a larger batch of 16 images.  Due to the memory requirements of this network, a single V100 instance can accommodate only batch size 1.  These networks were trained in parallel on 4 V100 devices, using gradient accumulation to emulate larger batch sizes.  Figure~\ref{fig:training_metrics} shows the progression of the metrics while training the baseline model.

\begin{table}
\centering
\begin{tabular}{rcccc}
\multicolumn{1}{l}{} & Acc. Non 0 & Cosmic IoU & Neutrino IoU & Mean IoU \\
Baseline             & 0.951      & 0.908      & 0.606        & 0.757    \\
Concat. Connections  & 0.947      & 0.898      & 0.609        & 0.753    \\
Cross-plane Blocked  & 0.942      & 0.898      & 0.571        & 0.734    \\
Batch Size x 2       & 0.956      & 0.914      & 0.698        & 0.806    \\
Convolution Upsample & 0.938      & 0.898      & 0.539        & 0.718    \\
Num. Filters / 2     & 0.930      & 0.887      & 0.457        & 0.672    \\
No Loss Balance      & 0.913      & 0.882      & 0.544        & 0.713    \\
Larger Learning Rate & 0.896      & 0.852      & 0.447        & 0.649    \\
Non Residual         & 0.944      & 0.904      & 0.584        & 0.744    \\
Adam Optimizer       & 0.904      & 0.852      & 0.509        & 0.680    \\
Full Balance         & 0.940      & 0.720      & 0.339        & 0.530
\end{tabular}
\caption{A comparison of the performance metrics for the various networks trained.  The best result in each metric is highlighted.  The ``Mean IoU" is the mean of the cosmic and neutrino IoU values. ``Acc. Non 0'' refers to the non-background accuracy.\label{tab:metrics} }
\end{table}

In Table~\ref{tab:metrics}, we compare the metrics for the different loss schemes and for the network with the concatenate operation blocked.  We see good performance in the baseline model, however the models with fully balanced loss and without a concatenate operation are degraded.  The full loss balancing exhibits a `blurring' effect around the cosmic and neutrino pixels, since the penalty for over-predicting in the vicinity of those points is minimal.  Since nearly half of all events have some overlap between cosmic and neutrino particles, this significantly degrades performance.  We also see that using a less extreme loss weighting performs better than no weighting at all, due to the relatively low number of neutrino pixels.  Notably, the network with the concatenate connections blocked at the deepest layer (therefore, no cross plane correlation), performs more poorly than the baseline model with every other parameter held constant.  Notably, the larger learning rate and use of the adaptive Adam optimizer give poor results with this network.

The larger batch size shows the best performance, including in the average of both IoU metrics.  The cosmic IoU is higher than the neutrino IoU due to the difference in difficulty in these labels: many more cosmic pixels implies that errors of a few pixels have a small effect on the cosmic IoU, and a large detrimental effect on the neutrino IoU.  We speculate that increasing the batch size further will improve results and will investigate this further with the use of a massive computing system needed to accommodate this large network at a high batch size for training.

As a final comment on the training process, we note that this network is expensive to train and has challenging convergence properties.  This has limited the experiments performed on model and training hyperparameters.  We expect a future result to investigate hyperparameters in a systematic way.  
In the following section, we use the model trained with a minibatch size of 16, ‘Batch Size x2’, as it had the best performance on the test set.

%\section{Example Analysis on SBND Simulation}
\section{Analysis Results}
\label{sec:analysis}
\begin{figure}
    \centering
    \includegraphics[width=0.7\columnwidth]{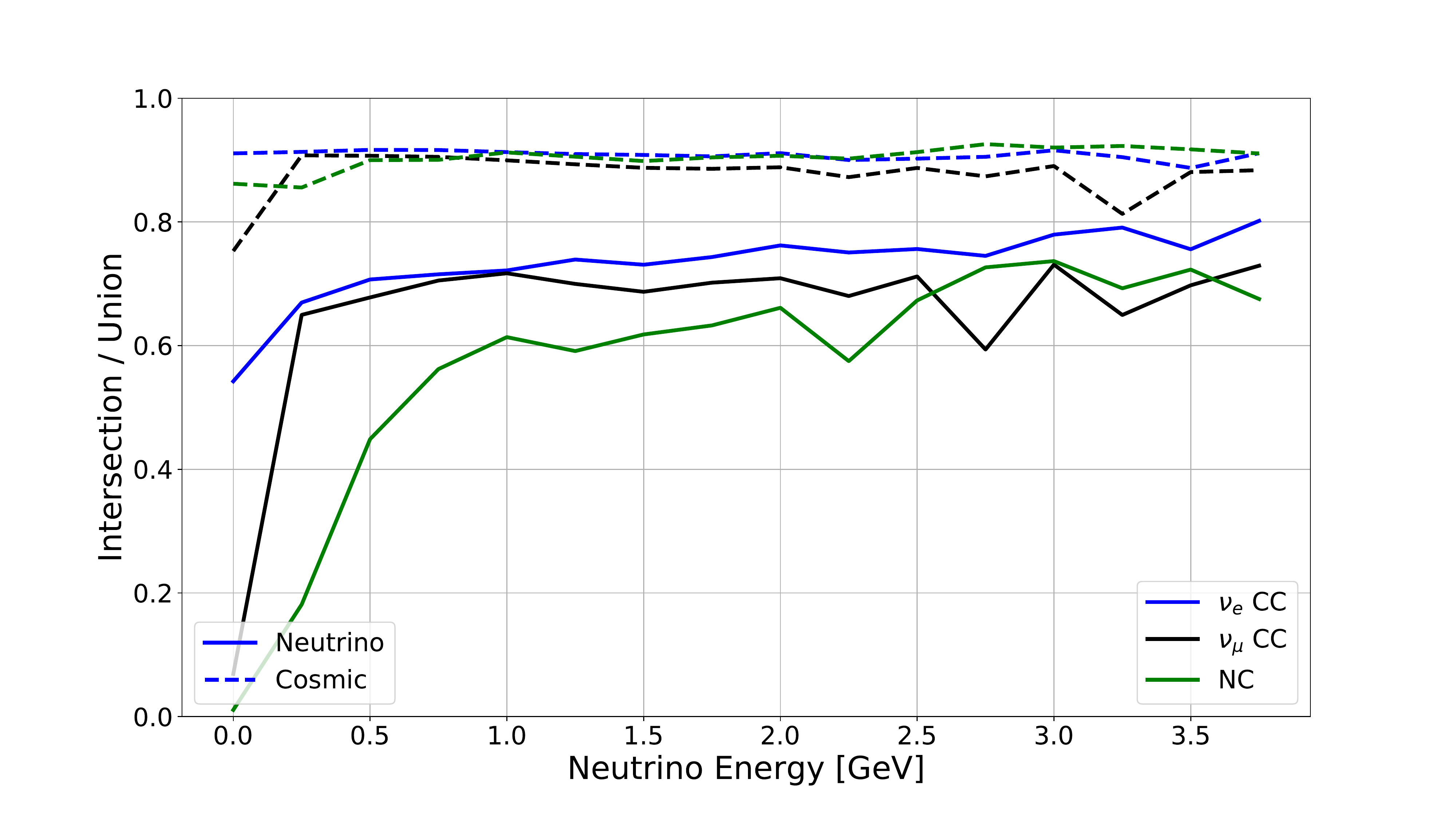}
    \caption{Metric performance across neutrino interaction types, as a function of neutrino energy.  The solid lines are the Intersection over Union for the neutrino predicted/labeled pixels, while the dashed lines are the Intersection over Union for the cosmic predicted/labeled pixels. Each color in this plot represents the IoU for all events containing that particular neutrino interaction.}
    \label{fig:bigger_batch_iou}
\end{figure}

Figure~\ref{fig:bigger_batch_iou} shows the metric performance as a function of neutrino energy for the best performing network, broken out across three kinds of neutrino interactions: electron neutrino charged current, muon neutrino charged current, and neutral current.

To demonstrate the utility of this deep neural network in a physics analysis, we perform a very elementary selection of events.  We perform inference on a selection of events from all types of simulated interactions, including events where there is no neutrino interaction.  
% In all of the below results, we are using the results of the best trained network, which is the baseline network with a minibatch size of 16.

There are two main objectives of this analysis.  First, on an event by event basis, decide if there is a neutrino interaction present in the measured charge using TPC information only.  It is expected that any additional information from the light collection or cosmic ray tagging systems will further enhance these results.  Second, within an interaction that has been selected as a neutrino interaction, measure the accuracy with which the interaction has been selected from the cosmic backgrounds.

To demonstrate the performance in event-level identification, we apply a simple set of metrics. We require a minimum number of pixels, per image, to be classified as neutrino by the network. Additionally, since the drift direction (Y-axis) of all three images is shared in each event, we apply a matching criterion.  Specifically, we compute the mean Y location of all neutrino-tagged pixels in each plane, and we require that the difference in this mean location is small across all three planes.

Quantitatively, we find good results by requiring at least 100 neutrino-tagged pixels per plane, and a maximum separation of mean Y location of 50 pixels across all three combinations of images.  With these basic cuts, we observe the selection efficiencies of Table~\ref{tab:selection}.  We note that neither 100 pixels per plane, nor a separation distance of 50 pixels, is a well tuned cut.  For some analyses targeting low energy events in the Booster Neutrino Beam, these cuts would be too aggressive.  Instead, the desired goal is to demonstrate that the predictive power of this network can be leveraged in a basic event filtering workflow.

\begin{table}
\centering
\begin{tabular}{rl}
Category & Efficiency \\
$\nu_e$ CC    & 91.5\%     \\
$\nu_\mu$ CC   & 78.6\%     \\
NC       & 37.3\%     \\
Cosmics  & 91.1\% cosmic-only event rejection
\end{tabular}
\caption{Selection efficiencies for sample cuts using the inference output of the best network.}
\label{tab:selection}
\end{table}
The selection efficiencies with these cuts, though not aggressively tuned, do have variation from one type of neutrino interaction to another.  The muon-neutrino events are distinguished by the presence of a long muon from the neutrino interaction, while electron neutrino events have no muons and instead an electro-magnetic shower.  Since the cosmic particles are primarily, though not entirely, composed of high energy muons, it is not surprising that electron neutrino events are more easily distinguished from cosmic-only events, as compared to muon neutrino events.  Additionally, the neutral current events have an outgoing neutrino that carries away some fraction of the energy of the event; on average, these events have much less energy in the TPC and therefore fewer active pixels to use for selection and discrimination of events.  Consequently, neutral current events are harder to reject compared to charged current events.

We do not speculate here on final purity for an analysis of this kind on the BNB spectrum of neutrinos at SBND.  The final analysis will use both scintillation light and cosmic ray tagger information in addition to the TPC data.  However, it is notable that a simple analysis can reduce the cosmic-only interactions by a factor of 10x, and the remaining events have the correct pixels labeled at a 95\% non-background accuracy level.  We believe this is a promising technique for the SBN experiments.

\section{Conclusions}

In this paper, we have demonstrated a novel technique for pixel level segmentation to remove cosmic backgrounds from LArTPC images.  We have shown how different deep neural networks can be designed and trained for this task, and presented metrics that can be used to select the best versions.
The technique developed is applicable to other LArTPC detectors running at surface level, such as MicroBooNE, ICARUS and ProtoDUNE.
We anticipate future publications studying the hyperparameters of these networks, and an updated dataset with a more realistic detector simulation prior to the application of this technique to real neutrino data.

\section{Acknowledgements}

%This research used resources of the Argonne Leadership Computing Facility, which is a DOE Office of Science User Facility supported under Contract DE-AC02-06CH11357.

The SBND Collaboration acknowledges the generous support of the following organizations: the U.S. Department of Energy, Office of Science, Office of High Energy Physics; the U.S. National Science Foundation; the Science and Technology Facilities Council (STFC), part of United Kingdom Research and Innovation, and The Royal Society of the United Kingdom; the Swiss National Science Foundation; the Spanish Ministerio de Ciencia e Innovación (PID2019-104676GB-C32) and Junta de Andalucía (SOMM17/6104/UGR, P18-FR-4314) FEDER Funds; and the S\~{a}o Paulo Research Foundation (FAPESP) and the National Council of Scientific and Technological Development (CNPq) of Brazil. We acknowledge Los Alamos National Laboratory for LDRD funding. This research used resources of the Argonne Leadership Computing Facility, which is a DOE Office of Science User Facility supported under Contract DE-AC02-06CH11357.  SBND is an experiment at the Fermi National Accelerator Laboratory (Fermilab), a U.S. Department of Energy, Office of Science, HEP User Facility. Fermilab is managed by Fermi Research Alliance, LLC (FRA), acting under Contract No. DE-AC02-07CH11359. 

\bibliographystyle{unsrt}
\bibliography{references}

\end{document}